\documentclass[%
 aip,
 jmp,%
 amsmath,amssymb,
preprint,%
]{revtex4-1}
\usepackage{graphicx}
\usepackage{epstopdf}
\usepackage{amsmath,amssymb}

\begin{document}
\title{Characterization of Traps at Nitrided SiO$_2$/SiC Interfaces near the Conduction Band Edge by using Hall Effect Measurements}

\author{Tetsuo Hatakeyama}
\affiliation{Advanced Power Electronics Research Center, 
National Institute of Advanced Industrial Science and Technology, 16-1 Onogawa, Tsukuba, Ibaraki 305-8569, Japan}
\email{tetsuo-hatakeyama@aist.go.jp}
\author{Yuji Kiuchi}
\author{Mitsuru Sometan}
\author{Shinsuke Harada}
\affiliation{Advanced Power Electronics Research Center, 
National Institute of Advanced Industrial Science and Technology, 16-1 Onogawa, Tsukuba, Ibaraki 305-8569, Japan}
\author{Dai Okamoto}
\author{Hiroshi Yano}
\affiliation{Graduate School of Pure and Applied Sciences, University of Tsukuba, 1-1-1 Tennodai, Tsukuba, Ibaraki, 305-8573, Japan}

\author{Yoshiyuki Yonezawa}
\author{Hajime Okumura}
\affiliation{Advanced Power Electronics Research Center, 
National Institute of Advanced Industrial Science and Technology, 16-1 Onogawa, Tsukuba, Ibaraki 305-8569, Japan}


\begin{abstract}
The effects of nitridation on the density of traps at SiO$_2$/SiC interfaces near the conduction band edge
were qualitatively examined by a simple, newly developed characterization method that utilizes Hall effect
measurements and split capacitance-voltage measurements. The results showed a significant reduction 
in the density of interface traps near the conduction band edge by nitridation, as well as the high density 
of interface traps that was not eliminated by nitridation.
\end{abstract}

\maketitle
\newcommand{\Ec}{E_\mathrm{C}}
\newcommand{\Ecs}{E_\mathrm{Cs}}
\newcommand{\Dit}{D_\mathrm{it}}
\newcommand{\Na}{N_\mathrm{A}}

\newcommand{\nfree}{n_\mathrm{free}}
\newcommand{\ntrap}{n_\mathrm{trap}}
\newcommand{\ntotal}{n_\mathrm{total}}
\newcommand{\Vg}{V_\mathrm{g}}
\newcommand{\Vfb}{V_\mathrm{FB}}
\newcommand{\Cgc}{C_\mathrm{gc}}
\newcommand{\Vmin}{V_\mathrm{min}}
\newcommand{\phiB}{\phi_\mathrm{B}}
\newcommand{\phis}{\phi_\mathrm{s}}
\newcommand{\psis}{\psi_\mathrm{s}}
\newcommand{\epss}{\varepsilon_\mathrm{s}}
\newcommand{\epsv}{\varepsilon_\mathrm{0}}
\newcommand{\kb}{k_\mathrm{B}}
\newcommand{\Nc}{N_\mathrm{C}}
\newcommand{\ndep}{n_\mathrm{dep}}
\newcommand{\Efs}{E_\mathrm{Fs}}
\newcommand{\FDF}{\mathcal F}
\newcommand{\muhall}{\mu_\mathrm{H}}
\newcommand{\mufe}{\mu_\mathrm{FE}}
\newcommand{\Et}{E_\mathrm{t}}
In recent years, silicon carbide (SiC) metal-oxide semiconductor field-effect transistors (MOSFETs) have been commercialized as the next-generation, high-voltage, low-loss power devices \cite{baliga}.
However, even in the case of state-of-the-art SiC MOSFETs, a high MOS channel resistance caused by low mobility in the SiO$_2$/SiC interfaces
still limits their potential performance. 
Therefore, the improvement of mobility in SiO$_2$/SiC interfaces remains a critical issue in the research and development of SiC MOSFETs. 
It was presumed that the high density of traps at the SiO$_2$/SiC interfaces near the conduction band edge ($\Ec$) was the cause of degradation in mobility \cite{afanasiev_pss,saks2000}. 
It is therefore highly essential to acquire the fundamental knowledge on the density of traps at the interface ($\Dit$) near $\Ec$, 
which will provide us with clues to further our understanding as to what the interface  traps are and how to reduce them.
However, it has been difficult to quantitatively examine the effect of $\Dit$ on the mobility of the fabricated SiC MOSFETs. 
This is partly because a well-known Hi-Lo capacitance-voltage ($C$-$V$) method is not suitable for the characterization of the energy distribution of the density of traps at the interface ($\Dit(E)$) near $\Ec$ \cite{yoshioka_psi}, and partly because of the lack of knowledge of the energy range of $\Dit$, which directly affects the mobility of SiC MOSFETs. In fact, $\Dit(E)$ characterized by a Hi-Lo $C$-$V$ method does not always correlate with the mobility \cite{Suzukiwet,Suzukino}. Unfortunately, a Hi-Lo $C$-$V$ method is still commonly used for the screening of newly developed processes, for the improvement of a SiO$_2$/SiC interface\cite{rfacemos,bariummos}. \par
On the other hand, to expand the limit of the energy range of $\Dit(E)$ towards $\Ec$, many characterization techniques were investigated, such as deep-level transient spectroscopy, a conductance method, and so on \cite{pensl2008, yoshioka_psi,hatakeyamadlts}. However, it turned out that these techniques expand the characterizable energy range of $\Dit(E)$ to some extent, but they are not effective enough to characterize $\Dit(E)$ near $\Ec$. Consequently, we formulated a new simple and practical characterization method for $\Dit(E)$ near $\Ec$, by using a split $C$-$V$ method and Hall effect measurements. In this letter, we first derive the formulae for this characterization method. Subsequently, this characterization method is applied to determine
 the effect of nitridation on $\Dit(E)$ near $\Ec$. It should be noted that the nitridation is the mainstream of the passivation process of traps at the SiO$_2$/SiC interface \cite{Rozen}.\par
The core idea of the new characterization method of $\Dit(E)$ near $\Ec$ is to focus on the quantitative characterization of the densities of free (mobile) and trapped carriers at SiO$_2$/SiC interfaces \cite{arnold2001,saks_dit,ortiz2015,takagi2014,tsujimura,hatakeyama2015,shiomimob}. The physics governing the trapping of electrons at the interface is included in these two characterized quantities. However, to extract the quantitative information on $\Dit(E)$ from them, a careful examination of the relationship between the applied gate voltage ($\Vg$) and the energy of the traps at the interface ($\Et$) is required \cite{hatakeyama2015,takagissdm}. In the new characterization method, the errors in the relationship between the $\Vg$ and the $\Et$ are minimized by the use of the analytic formulae derived from the exact solution of the Poisson equation in a MOS structure \cite{nicollian_ns}. \par
Now, the new characterization method of $\Dit(E)$ near $\Ec$ is described step by step. In the first step, the free carrier density ($\nfree(\Vg)$) is characterized as a function of $\Vg$, by the use of Hall effect measurements for a SiC MOSFET. In the second step, the capacitance between the gate and the channel ($\Cgc(\Vg)$) of a SiC MOSFET is characterized as a function of a $\Vg$, by the use of a split $C$-$V$ technique. In the third step, the total density of carriers at the interface, induced by the applied $\Vg$ ($\ntotal(\Vg)$), is obtained by integrating the $\Cgc(V)$ over the scanned voltages ($V$) as follows:
\begin{equation}
\ntotal(\Vg) = \int^{\Vg}_{\Vmin} \Cgc(V) dV,
\label{ntotalcal}
\end{equation}
where $\Vmin$ is the minimum value of the scanned gate voltage, which is set to be less than the threshold voltage. Finally, the trapped carrier density ($\ntrap(\Vg))$ is obtained by subtracting $\nfree(\Vg)$ from $\ntotal(\Vg)$, as follows:
\begin{equation}
\ntrap(\Vg) =\ntotal(\Vg)-\nfree(\Vg).
\label{ntrapcal}
\end{equation}
From these obtained functions ($\ntrap(\Vg)$ and $\nfree(\Vg)$), $\Dit(E)$ near $\Ec$ is calculated according to the following formulae.\par
The density of the trapped electrons ($\ntrap$) as a function of the Fermi level at the interface ($\Efs$) is given by the product of the density of interface traps $\Dit(E)$ and the Fermi--Dirac distribution function, as follows:
\begin{equation}
\ntrap(\Efs)=\int_{-\infty}^{\infty} \Dit(E) \frac {dE}{1+g \exp \left[ \beta (E- \Efs) \right]}
\label{eq:ntrap}
\end{equation}
where $\beta$ and $g$ are the thermodynamic beta factor and the degeneracy factor. 
In this study,  $g$ is assumed to be 2. 
When we examine $\Dit(E)$ and $\Efs$ at the interface, 
the origin of the energy is implicitly set to the intrinsic Fermi energy of the SiC at the interface. 
In this notation, the $\Efs$ changes as $\Vg$ changes. In the standard energy diagram of a MOS interface, $\Efs$ is essentially the same as the surface potential at the interface $\phis$, as follows:
\begin{equation}
\Efs=e \phis.
\end{equation}
The derivative of Eq. \ref {eq:ntrap} gives the following relationship by approximating the derivative of the Fermi--Dirac distribution function to the Dirac delta function, as follows:
\begin{equation}
\frac{d \ntrap}{d \Efs} \approx \Dit \left( \Efs -\beta^{-1} \ln g \right),
\end{equation}
where the shift in energy in the argument of $\Dit(E)$ comes from the degeneracy factor $g$. Thus, $\Dit(E)$ is given by the derivative of $\ntrap(\Vg)$ with respect to the $\Efs$. The relationship between the trap energy ($\Et$) (which is the argument of $\Dit(E)$) and $\Efs$, is given by the following formula:
\begin{equation}
\Et=\Efs - \beta^{-1} \ln g.
\end{equation}
It should be noted that the measured trapped carrier density is obtained as a function of $\Vg$, and not $\Efs$. Thereby, a parametric differentiation is introduced as follows:
\begin{equation}
\frac{d \ntrap}{d \Efs} =\frac{ \cfrac{d \ntrap(\Vg)}{d \Vg}   } {\cfrac{d \Efs(\Vg) }{d \Vg} }.
\end{equation}
The numerator of this equation was obtained by the numerical derivative of $\ntrap(\Vg)$.
\begin{table}[tbp]
\begin{center}
\caption{%
Abbreviated names, post-oxidation-annealing in nitric oxide (NOPOA) conditions (the duration time of NOPOA) and 
doping densities of the p-well of the prepared SiC MOSFETs}\label{tablepoa}
\begin{tabular} {c c c} \hline
name & NO POA (min.) & $N_A-N_D$ (cm$^{-3}$) \\ \hline \hline
Dry  & 0  & $2.4\times 10^{15}$ \\ \hline
NO10 & 10 & $1.3\times 10^{15}$\\ \hline
NO60 & 60& $2.3\times 10^{15}$ \\ \hline
NO120 & 120 & $2.7\times 10^{15}$ \\ \hline
\end{tabular}
\end{center}
\end{table}
\begin{figure}[tbp]
\begin{center}
\includegraphics[width=8cm]{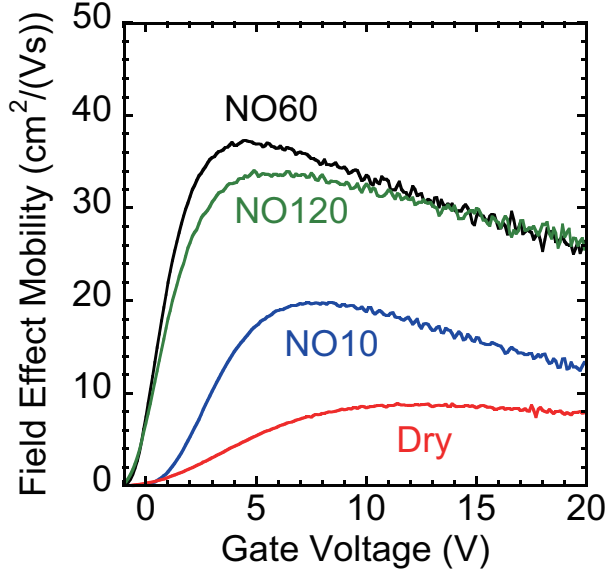}
\end{center}
\caption{%
Field effect mobilities ($\mufe$'s) of the four prepared types of SiC MOSFETs on Si-face (Dry, NO10, NO60, and NO120)
}
\label{fmob}
\end{figure}
The denominator of the equation was calculated by using $\nfree(\Vg)$, as described below. Under Fermi-Dirac statistics, the $\nfree$ in strong inversion is approximated by a function of the $\Efs$, as follows \cite{nicollian_ns}:
\begin{equation}
\begin{split}
&\nfree  \cong   \sqrt{2 e^{-2}\epss \epsv \kb T} \\ 
& \times \left[ \Na \beta e \psis  + \Nc \FDF_{3/2} \left \{ -\beta (\Ecs - \Efs)  \right \}  \right] ^{1/2} - \ndep
\end{split}
\label{eq:fdnfree}
\end{equation}
where $\epss$, $\epsv$, $\kb$, $T$, $\Nc$, $\Na$, and $\psis$ are the relative permittivity of SiC, vacuum permittivity, Boltzmann constant, temperature, effective density of states in the conduction band, acceptor density in the p-well, and band bending at the interface, respectively. $\Ecs$ is the energy at the bottom of the conduction band at the interface. $\FDF_{3/2}(x) $ is a complete Fermi-Dirac integral for an index of $3/2$. The $\ndep$ is the areal acceptor density in the depletion layer in a strong inversion state of a SiC MOSFET, which is approximated by the following equation:
\begin{equation}
\ndep \approx \sqrt{4  e^{-1} \epss \epsv \Na \phiB},
\end{equation}
where $\phiB$ is the bulk potential. The $\Efs$ as a function of $\nfree$ is derived from Eq. \ref{eq:fdnfree} as follows:
\begin{equation}
\Efs-\Ecs \cong \kb T \FDF_{3/2}^{\: -1} \left [ \frac{e^2(\nfree^2+2\nfree \ndep)}{2\epss \epsv \kb T \Nc}   \right]
\label{eq:FDEfs}
\end{equation}
where $\FDF_{3/2}^{\: -1}(x)$ is an inverse function of $\FDF_{3/2}(x)$. The differential equation of $\Efs$ with respect to $\Vg$ can be obtained by the parametric differentiation as follows:
\begin{equation}
\frac{d \Efs}{d \Vg} \cong 2\kb T \frac{\nfree+\ndep}{ \epss \epsv\Nc \FDF_{1/2}[-\beta (\Ecs-\Efs)]}
\frac{d \nfree}{d \Vg},
\label{eq:FDdEdVg}
\end{equation}
where the differential of $\nfree$ with respect to $\Vg$ in the right hand side of the equation can be obtained by the numerical differentiation of the measured $\nfree(\Vg)$ value. \par
In the evaluation of Eq. \ref{eq:FDEfs} and Eq. \ref{eq:FDdEdVg}, analytical approximations of $\FDF_{j}(x)$ are used \cite{blackmore_fd}.\par
The characterization method described above was applied to determine the effect of nitridation on $\Dit(E)$, in SiO$_2$/SiC interfaces.
\begin{figure}[tbp]
\begin{center}
\includegraphics[width=15.0cm]{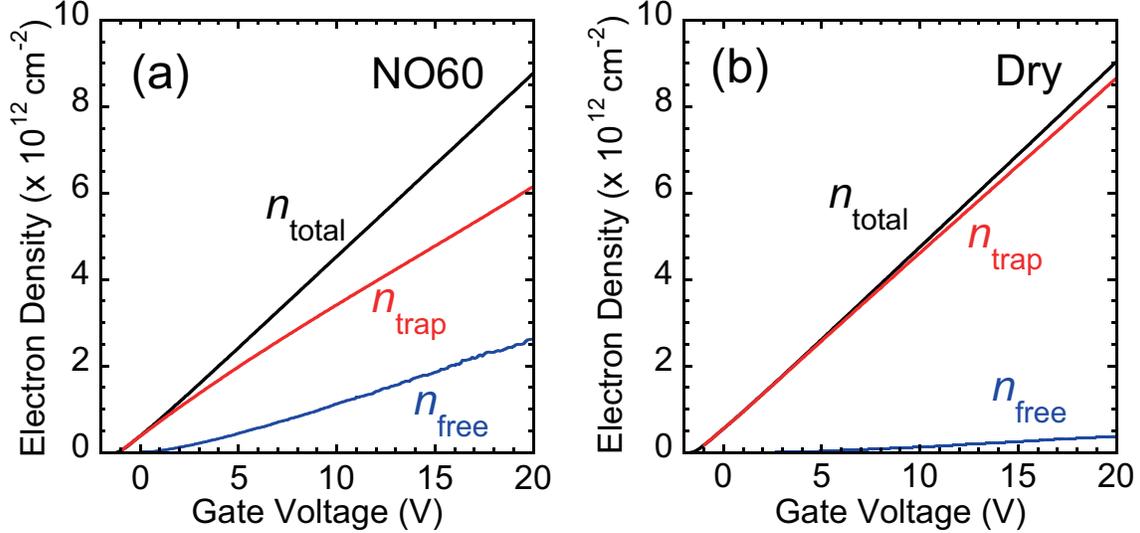}
\end{center}
\caption{%
$\ntotal (\Vg)$, $\ntrap(\Vg)$, and $\nfree(\Vg)$ of (a) an ``NO60'' SiC MOSFET and (b) a ``Dry'' SiC MOSFET.
}
\label{ntrap}
\end{figure}
Four types of SiC MOSFETs on Si-faces (4H-SiC (0001)) were prepared and characterized. The thickness of the thermally grown gate oxide was around 50 nm. Post-oxidation annealing in nitric oxide (NO POA) at $1250 \ {}^\circ\mathrm{C}$ was applied to three of the four samples. The abbreviated names, process conditions of NO POA, and doping densities of the p-well of SiC MOSFETs, are summarized in table \ref{tablepoa}. The gate electrode was phosphorus doped poly-silicon. \par
The Hall effect measurements, using the van der Pauw technique, were carried out to obtain $\nfree(\Vg)$ and the Hall mobility ($\muhall$) of the prepared SiC MOSFETs. The strength of the applied magnetic field is about 0.5 T. The Hall factor ($\gamma$) is assumed to be 1.\par
The total carrier density ($\ntotal(\Vg)$) was calculated by integrating the $\Cgc$, which was obtained by using the split-$C$-$V$ technique with respect to the gate voltage, according to Eq. \ref{ntotalcal}. The split-$C$-$V$ measurements were conducted at 10 Hz, using the ultra-low frequency $C$-$V$ measurement system created by Agilent Technologies. In the measurements, the gate voltage was scanned from the inversion region to the depletion region, in order to suppress the variation of the measurements caused by the shift of the threshold voltage. We note that one of the advantages of using the split $C$-$V$ technique is that we can exclude the errors derived from the assumption of the threshold voltage, when calculating $\ntotal$. The trapped carrier density ($\ntrap(\Vg)$) was obtained according to Eq. \ref{ntrapcal}.\par

The field effect mobilities ($\mufe$'s) of four types of prepared SiC MOSFETs are shown in Fig. \ref{fmob}. It can be seen that $\mufe$ was improved by NO POA for at least 60 min. However, the effect of nitridation on $\mufe$ was saturated when the duration time of NO POA was more than 60 min. Furthermore, it should be noted that the observed maximum value of the $\mufe$ of the best sample of ``NO 60'' is less than 40 cm$^2/$(Vs). This value is much smaller than the bulk mobility of 4H-SiC, which is around 1000 cm$^2/$(Vs) \cite{hatakeyamamob}. By the examination of the $\mufe$'s, the issues concerning the electron transport of the nitrided SiO$_2$/SiC interfaces are summarized as follows:
(1) the mechanism of improvement and the saturation of $\mufe$ by nitridation,
(2) the reason why $\mufe$ of the nitrided SiO$_2$/SiC interfaces is much smaller than the bulk mobility of 4H-SiC. Hereafter, we discuss these issues based on the results obtained using our new characterization method.\par
\begin{figure}[tbp]
\begin{center}
\includegraphics[width=8cm]{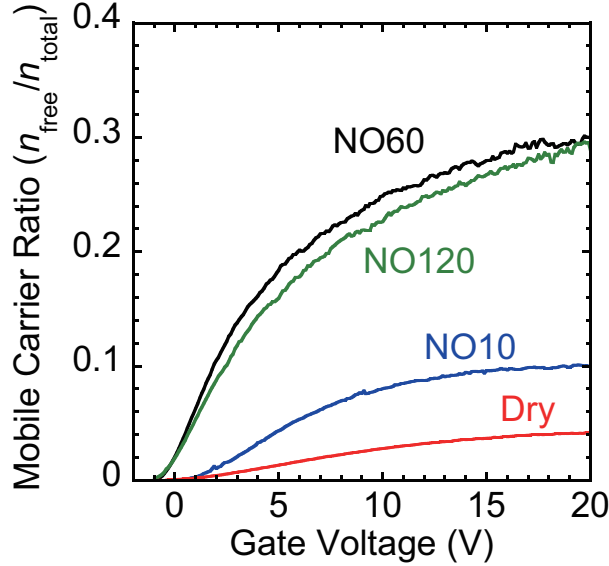}
\end{center}
\caption{%
Comparison of the ratios of $\nfree(\Vg)$ to $\ntotal(\Vg)$ (free carrier ratios) of 
four types of SiC MOSFETs on Si-face (Dry, NO10, NO60, NO120)}
\label{nsratio}
\end{figure}
In Fig. \ref{ntrap} (a) and (b), the obtained $\ntotal(\Vg)$, $\nfree(\Vg)$ and $\ntrap(\Vg)$ of the ``NO60'' and ``Dry'' samples are shown. In both cases, the trapping of electrons at the interface traps is dominant. Only a portion of the electrons induced by the gate voltage contribute to the transport in SiC MOSFETs, even in the case of the ``NO60'' sample. To quantitatively examine the effect of trapping on the electron transport, the ratios of $\nfree(\Vg)$ to $\ntotal(\Vg)$ (free carrier ratios) of the four prepared samples are shown in Fig. \ref{nsratio}. As shown in Fig. \ref{nsratio}, the free carrier ratios increased as the gate voltage was increased. The maximums of the free carrier ratios of ``Dry,'' ``NO10,'' ``NO60,'' and ``NO120,'' were 4\%, 10\%, 30\%, and 30\%, respectively. The free carrier ratio increased as the duration time of NO POA was increased up to 60 min. This relationship between the free carrier ratio and the duration time of NO POA is the same as
  that of the field effect mobility. Furthermore, the small free carrier ratio also explains, at least to some extent, the reason why the mobility of the nitrided SiO$_2$/SiC interfaces is much smaller than the bulk mobility of 4H-SiC.\par
\begin{figure}[tbp]
\begin{center}
\includegraphics[width=8cm]{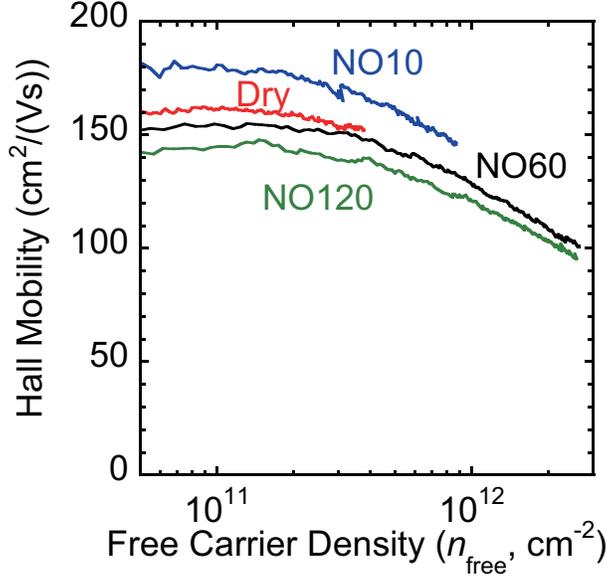}
\end{center}
\caption{%
Comparison of Hall mobilities ($\muhall$'s) as a function of the free carrier density of 
SiC MOSFETs on Si-face (Dry, NO10, NO60, NO120)
}%
\label{hmobns}
\end{figure}
To examine the effect of nitridation on the mobility of free carriers, the $\muhall$'s of the prepared samples are shown in Fig. \ref{hmobns} as a function of $\nfree$. All the $\muhall$'s gradually decreased as the free carrier densities increased. A distinct correlation between the $\muhall$ and the duration time of NO POA was not observed, whereas a weak correlation between the $\muhall$ and the doping density of the p-well of the SiC MOSFET was observed. From Figs. \ref{nsratio} and \ref{hmobns}, we conclude that the increase of the $\mufe$ by nitridation is caused by the increase of the free carrier ratio, 
and not by the increase of the $\muhall$ mobility. Furthermore, from the results obtained from Fig. \ref{hmobns}, we can see that the $\muhall$ was not sensitive to the reduction of $\ntrap$ caused by nitridation. We note that the trapped carriers at the interface become the Coulomb scattering centers in electron transport. Accordingly, the obtained results show that the density of the 
 Coulomb scattering centers at the interface have a limited effect on the $\muhall$. Therefore, we infer that the dominant scattering mechanism of the $\muhall$ is not the remote Coulomb scattering from the trapped carriers. \par
\begin{figure}[tb]
\begin{center}
\includegraphics[width=8cm]{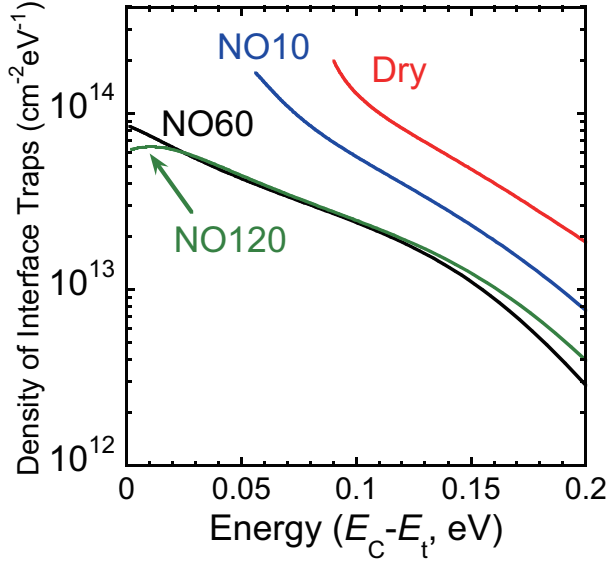}
\end{center}
\caption{%
Comparison of $\Dit(E)$ near $\Ec$ among prepared
samples (Dry,  NO10,  NO60, NO120) 
}%
\label{Dit}
\end{figure}
Finally, in Fig. \ref{Dit}, we show the $\Dit(E)$'s near $\Ec$, which were calculated from the measured $\ntrap(\Vg)$'s and $\nfree(\Vg)$'s of four samples, according to the new characterization method. In Fig. \ref{Dit}, all the $\Dit(E)$'s are more than 10$^{13}$ cm$^{-2}$eV$^{-1}$ around $\Ec$. We note that the possible errors derived from the non-unity value of the Hall factor are negligible in this case\cite{takagissdm}. Compared to the $\Dit(E)$ of the``Dry'' sample, the $\Dit(E)$'s of the nitrided samples, are reduced down to 25 \% at $\Ec-\Et =0.1$ (eV), particularly those of ``NO60'' and ``NO120.'' However, they are still more than $10^{13}$ cm$^{-2}$eV$^{-1}$ around $\Ec$. This non-reduced $\Dit(E)$ near $\Ec$ leads to the relatively low $\mufe$'s of the fully nitrided ``NO60'' and ``NO120'' samples, which are shown in Fig. \ref{fmob}. From these results, we conclude that nitridation reduces $\Dit(E)$ near $\Ec$, but is unable to completely eliminate them. \par
In summary, a simple, practical, and quantitative characterization method for $\Dit(E)$ near $\Ec$ was formulated by utilizing the split $C$-$V$ technique and Hall effect measurements. This characterization method was applied to the nitrided SiO$_2$/SiC (0001) interfaces 
to determine the effects of nitridation on $\Dit(E)$ and electron transport at the interfaces.
 It is concluded that nitridation reduces $\Dit(E)$ near $\Ec$, but cannot eliminate them completely. It is also concluded that $\muhall$ is not sensitive to the change of $\Dit(E)$ caused by nitridation. It is inferred that the effect of Coulomb scattering from the trapped electrons on the $\muhall$ may be limited.
\begin{acknowledgments}
This work was supported by the Council for Science, Technology and Innovation (CSTI), Cross-ministerial Strategic Innovation Promotion Program (SIP), ``Next-generation power electronics'' (funding agency: NEDO).
\end{acknowledgments}

\end{document}